\documentclass[10pt,journal,compsoc]{IEEEtran}
\ifCLASSOPTIONcompsoc
  \usepackage[nocompress]{cite}
\else
  \usepackage{cite}
\fi
\ifCLASSINFOpdf
\usepackage[pdftex]{graphicx}
\else
\fi
\usepackage{array}
\begin{document}
\title{\title{Development of Rehabilitation System (ReHabgame) through Monte-Carlo Tree Search Algorithm}}
\author{Shabnam~Sadeghi~Esfahlani,~\IEEEmembership{Fellow,~IET,}~George~Wilson,~\IEEEmembership{Member,~IEEE},
        % <-this % stops a space
\IEEEcompsocitemizethanks{\IEEEcompsocthanksitem S. S. Esfahlani and G. Wilson are with the Department of Computing and Technology, Anglia Ruskin University, Cambridge, UK \protect\\
% note need leading \protect in front of \\ to get a newline within \thanks as
% \\ is fragile and will error, could use \hfil\break instead.
shabnam.sadeghi-esfahlani@anglia.ac.uk, george.wilson@anglia.ac.uk}}

\IEEEtitleabstractindextext{
\begin{abstract}
Computational Intelligence (CI) in computer games plays an important role that could simulate various aspects of real life problems. CI in real-time decision-making games can provide a platform for the examination of tree search algorithms. In this paper, we present a rehabilitation serious game (ReHabgame) in which the Monte-Carlo Tree Search (MCTS) algorithm is utilized. The game is designed to combat the physical impairment of post stroke/brain injury casualties in order to improve upper limb movement. Through the process of ReHabgame the player chooses paths via upper limb according to his/her movement ability to reach virtual goal objects. The system adjusts the difficulty level of the game based on the player's quality of activity through MCTS. It learns from the movements made by a player and generates further subsequent objects for collection. The system collects orientation, muscle and joint activity data and utilizes them to make decisions. Players data are collected through Kinect Xbox One and Myo Armband.  The results show the effectiveness of the MCTS in the ReHabgame that progresses from highly achievable paths to the less achievable ones, thus configuring and personalizing the rehabilitation process.
\end{abstract}

\begin{IEEEkeywords}
Monte Carlo Tree Search, Gesture Recognition, Myo Armband, Kinect XBox One, Motion Capture.
\end{IEEEkeywords}}

% make the title area
\maketitle

\IEEEdisplaynontitleabstractindextext

\IEEEpeerreviewmaketitle

\IEEEraisesectionheading{\section{Introduction}\label{sec:introduction}}

\IEEEPARstart{R}{ecent} advances in low cost depth sensing technology have led to the creation of consumer electronics devices that can sense the user$'$s motion \cite{holden2002virtual}. These devices have rapidly become a major source of motivation for researchers who immediately recognized the potential use of the new technology in experimental rehabilitation \cite{holden2005virtual},\cite{vien2013monte}. It transforms the player into the controller, making games more intuitive to play, thus more accessible to a broader audience. Human-computer interaction systems such as {“Microsoft Kinect in 2010\footnote{http://www.xbox.com/en-US/xbox-one/accessories/kinect} and Thalmic Labs Myo in 2016\footnote{https://www.myo.com/}” have been utilised in scientific research projects and commercial applications and are shown in Fig.~\ref{Fig. 1}. These sensors capture player movements in the real world/time and convey them inside the game as well as capturing the position, orientation, and muscle activities. Kinect depth and image sensors facilitate a robust interactive human body tracking facility which reads the environment through a randomized decision forest algorithm. Its single input depth image is segmented into a dense probabilistic body part labeling, with the parts defined being spatially localized near skeletal joints of interest \cite{shotton2013real}. A wireless Myo armband sensor provides data through eight electromyography (EMG) pods and a nine-axis inertial measurement unit (IMU) capturing real world movements through the use of a gyroscope and accelerometer. \\
A clinical application of a body based control game via sensor input is not a one-size-fits-all solution and needs to be customized according to the individual's condition, needs, and range of ability \cite{suma2013adapting}. The rehabilitation therapy requires real-time decision making and must be able to adapt to the difficulty level according to the rehabilitation goals.
Artificial and computational intelligence have mostly focused on games where no information is hidden and all actions are deterministic \cite{lucas2015artificial}. However, rehabilitation games feature elements of hidden information, real-time decision making, and uncertainty which are non-deterministic. Hidden information occurs when a game is in one of the many possible states with uncertainties, and real-time decision making is required to balance the performance and gain score. \\
In this research, we propose the use of Monte-Carlo Tree Search (MCTS) in designing a serious game for rehabilitation purposes called RehabGame. This virtual reality therapy system exposes stroke or traumatic brain injury survivors to a 3D environment to monitor, measure and improve the kinematic movement with the focus on the upper limbs. It is done through motion capture sensors (Kinect and Myo) that are interfaced with a game engine (Unity3D) through plugins Microsoft SDK 2015\footnote{http://wiki.etc.cmu.edu/unity3d/index.php/Microsoft} and Kinect tools and resources 2016\footnote{https://developer.microsoft.com/en-us/windows/kinect/tools}. It measures the kinetic and kinematic movements through reach and grab which is a real time rehabilitation therapy.

\subsection{Background \& Related Work}
In the last few years, several Monte-Carlo based techniques have emerged in the field of computer games that have had significant success in various AI game playing algorithms \cite{lucas2015artificial}. They can be applied to a single-player games and planning problems, multi-player games, real-time games, and games with uncertainty or simultaneous moves. The MCTS method was developed by combining the random sampling of traditional Monte-Carlo methods with random simulations to search for the best outcome. Possible results are organized in a search tree to estimate the long-term potential, and with each iteration, the game tree is expanded and converged over time \cite{finnsson2012generalized}. The technique specifies which actions are possible from each state and which states are terminal \cite{whitehouse2014monte}. 
The approach is based on real-time planning that finds the best branch and plays the best arm within that branch. 
MCTS has been applied successfully to many board games including; the Asian board game GO \cite{billings2002challenge} and \cite{sheppard2002world}, General Game Playing where the rules of games used to evaluate techniques are not known in advance \cite{finnsson2012generalized} \cite{frydenberg2015investigating} and \cite{de2016monte}, imperfect information games where each player independently chooses an action and these actions are applied at the same time, such as Scrabble and Bridge \cite{browne2012survey}, the arcade Ms Pac-Man game with repeated random sampling to obtain results \cite{maycock2013enhancing}, \cite{nguyen2013monte}, \cite{tong2011monte}, \cite{pepels2014real} and \cite{ikehata2011monte}. 
\begin{figure}[thpb]
      \centering
      %\framebox{\parbox{1in}{}}
     \includegraphics[scale=0.6]{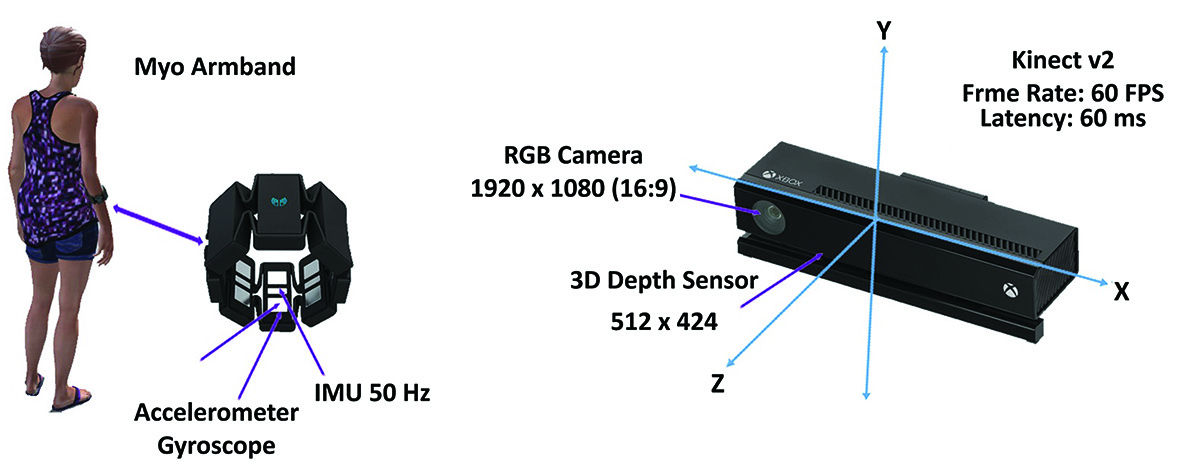}
 \caption{Main components of the Kinect and Myo device.}
      \label{Fig. 1}
   \end{figure}

In the last few years, several integrated solutions have been proposed for computer-based rehabilitation techniques \cite{carlozzi2013using}, \cite{mohanty2011teaching}, \cite{weiss2013video}, \cite{mirelman2009effects}, \cite{o2013cognitive}, \cite{ustinova2014virtual}, \cite{villiger2012virtual}, \cite{levin2015emergence}, \cite{villiger2011virtual}, \cite{saposnik2010effectiveness}, \cite{dobkin2004strategies}, \cite{tsekleves2016development}, \cite{smith1999task}, \cite{page2004efficacy} and \cite{holden2002virtual}. Although there has been a lot of research in the area of virtual reality there are still outstanding challenges to apply MCTS technique to rehabilitation games in order to adapt it to patient's status. \cite{suma2013adapting} developed a middleware software framework called the Articulated Skeleton Toolkit to integrate full-body interaction with VR and video games. They presented a system architecture and described two case studies to evaluate the toolkit for controlling the first person video games. A study by \cite{pirovano2012self} suggests that rehabilitation games should be integrated into general-purpose rehabilitation stations, adhere to the constraints posed by the clinical protocols, and adapt to patients status and progress. They used a Fuzzy system to monitor the execution of the exercises and provide direct feedback to the patients as well as a Quest Bayesian adaptive approach for real-time adaptation. \cite{frydenberg2015investigating} studied MCTS methods to be implemented in more complex video games. They investigated the use of a modified MCTS algorithm in video game playing using the general video game AI framework. In particular a modification to the core of MCTS known as UCT was performed and the results of their research showed the effectiveness and efficiency of the modified MCTS highly depends on the type of the game designed. \cite{pepels2014real} and \cite{nguyen2013monte} have introduced MCTS for controlling the characters in the real time game Ms Pac-Man to find an optimal path for an agent at each turn. It determines the best move to make, based on the results of numerous randomized simulations.  

\section{Kinect and Myo Devices}
The Kinect One $V$$2$ device consists of an infrared laser based IR emitter and a colored (RGB: red-green-blue) camera.  It can detect the position and orientation of $25$ individual joints (including thumbs), the weight put on each limb, a speed of joint movements, and track gestures performed with a standard controller. Body position is determoned in a two stage process; firstly computing a depth map and secondly infering participant body position. The depth map is constructed by analyzing a speckle pattern of infrared laser light \cite{criminisi2011decision}. Whilst body parts are inferred using a randomized decision forest learned from over one million training examples \cite{maccormick2011does}. Kinect provides approximately $60$ skeleton frames per second.\\ The Myo armband streams the accelerometer, gyroscope, orientation, and the electromyograph (EMG) at $200$ $Hz$ frequency and the inertial measurement unit (IMU) data at $50$ $Hz$. It is made of three medical grade stainless steel EMG sensors that detect the electric impulses in the muscles. The armband is connected via a Bluetooth USB adapter and controls muscle activity related to movements of fingers, thumb, hand, wrist, and forearm.  
\begin{figure}[thpb]
      \centering
      %\framebox{\parbox{1in}{}}
       \includegraphics[scale=0.7
     ]{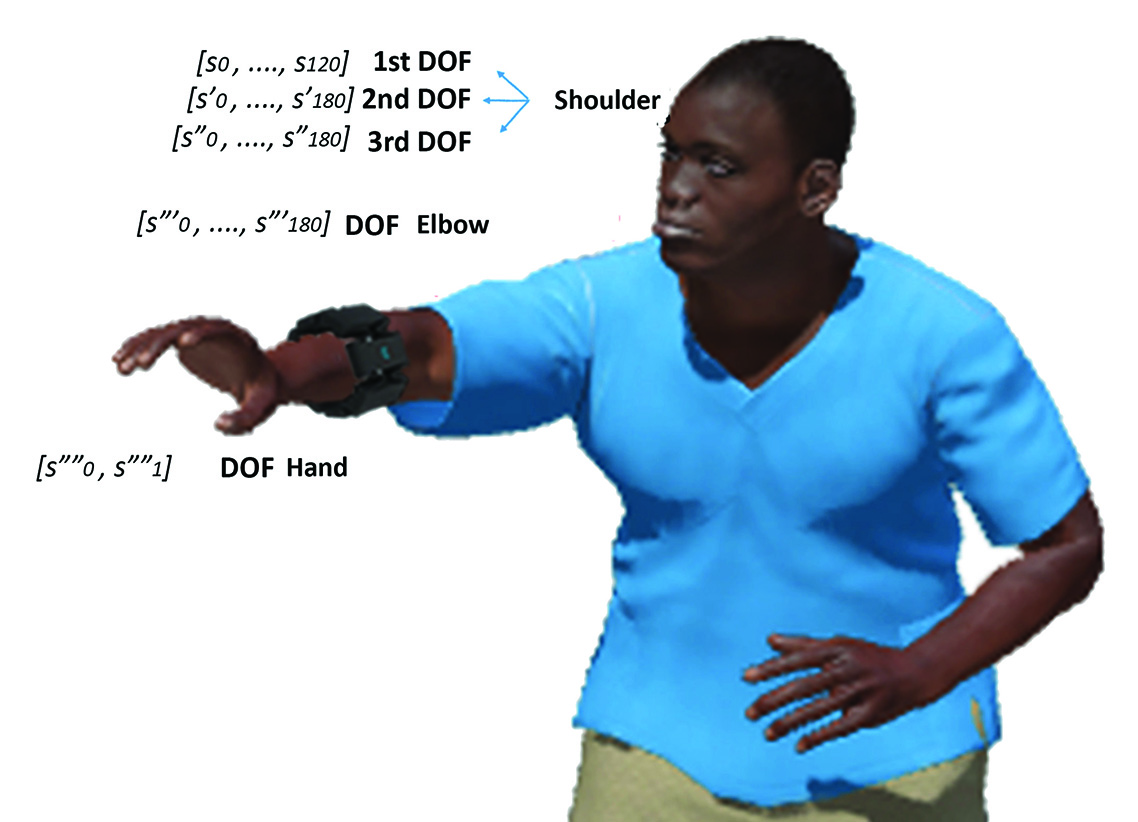}
      \caption{Character representation of the rehabilitation game with shoulder, elbow, and hand degree of freedom (DOF).}
      \label{Fig. 2}
   \end{figure}
Myo detects the electrical signal produced by muscle exertion through electromyography. It is used to discover underlying electrical activity displayed as a waveform (an electromyogram). A three-axis accelerometer, magnetometer, and gyroscope senses motion and records and collects real-time data with high accuracy and precision.

\section{Development of ReHabgame system}
A 3D character (avatar) is first constructed with 65 skeleton joints consists of one and three degree of freedom/s (DOF) developed in Mixamo\footnote{https://www.mixamo.com/fuse} $^,$ \footnote{https://www.assetstore.unity3d.com}. One degree of freedom joints include the elbow and knee, whilst spherical three DOF joints include the shoulder, wrist, ankle, and pelvis. Fig.~\ref{Fig. 2} shows the character with three DOF of shoulder, one DOF of elbow and three DOF of wrist.\\ 
In order to employ the MCTS framework for rehabilitation therapy four enhancements are defined \cite{ikehata2011monte} and \cite{pepels2014real} which are;

\begin{enumerate}[]
\item a constant tree depth with variable angles for the shoulder, elbow, and hand child elements in order to develop the parent upper limb tree search as shown in Fig.~\ref{Fig. 2} and Fig.~\ref{Fig. 4}.\\ 
Fig.~\ref{Fig. 2} for example illustrates shoulder rotation of $120 \deg$ in the x-direction, $180 \deg$ in the y-direction, $180 \deg$ in the z-direction, followed by rotation of the elbow by $90 \deg$, and finally hand movement with thumb and four fingers.  
\item search tree navigation and goal (virtual fruit) generation can be simulated according to the avatar's performance. 
\item long term participant goals include achievement of high scores, improved performance, and motivation. 
\item reuse of the search tree over several moves in a systematic way including backpropagation.   
 \end{enumerate}
The environment of the game is made of a garden, an avatar, various fruits that are rotating around their axis in the scene and awaiting the avatar to reach and collect them in the basket. Fruit spawning, participant grab and release time and position are registered. A timer is defined for each fruit with the clock starting as soon as it is generated. The tree is constructed from the path movements of the hand in each session.\\ 
Fruits need to be released at the top of the basket where the inner surface of it changes to green when it is released in a right location as shown in Fig.~\ref{Fig. 5}.  A warning is issued if the path is taken by a body and upper limb exceed constraints or the movement is outside the tree search. Position, angle, velocity and body constrains are specified as numerical values to minimize any risk of harm for users and at the same time encourage motor movement.\\
Although achieving high scores is the objective of the RehabGame the agent is designed based on player's performance and medical needs in order to take the player through a process that avoids frustration or harm. A randomness to fruit generation is introduced into the game to avoid pre-deterministic fixed paths during the game play. \\
The agent measures the distance between avatar to object utilizing a half arm span. Arm span or reach is the physical measurement of the length from one end of an individual's arms to the other when raised parallel to the ground at shoulder height at a $90  \deg$ angle \cite{tan2009arm}.
Distance vectors are accurately calculated including the shoulder to elbow $(P_1P_2)$ vector, elbow to wrist $(P_2P_3)$ vector, wrist to finger tips $(P_3P_4)$ vector as well as shoulder to objects $(P_1P_4)$ vector as illustrated in Fig.~\ref{Fig. 3}. 
\begin{figure}[thpb]
      \centering
     % \framebox{\parbox{100in}{}}
     \includegraphics[scale=0.8]{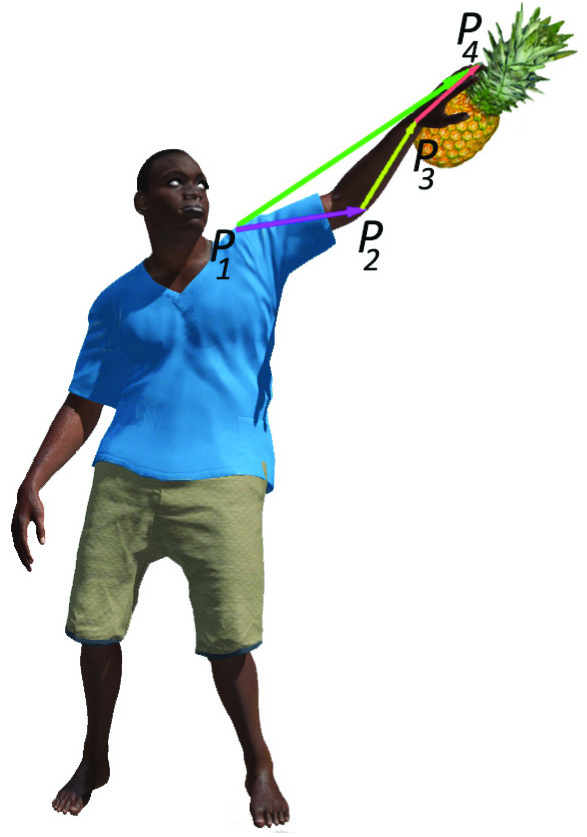}
      \caption{Distances between body joints and fruits is calculated by the agent.}
      \label{Fig. 3}
   \end{figure}

\section{Monte-Carlo Tree Search}
\subsection{General MCTS}
Monte-Carlo Tree Search is a probabilistic algorithm that uses lightweight random simulations to selectively grow a game tree. The method combines the precision of a tree search with the generality of random sampling \cite{browne2012survey}. The algorithm plans actions with a best first search technique that is based on stochastic models by Monte-Carlo simulations. In game play, this means that decisions are made based on the results of randomly simulated playouts \cite{pepels2014real}. Fig.~\ref{Fig. 4} shows the mechanism of the MCTS algorithm applied to a round of the RehabGame. 

Decisions in the tree are the movements of upper limb followed by traversing the tree during the selection step. It means that the upper limb rotates in accordance to the selected nodes until playout starts based on the random path through the tree. The tree represents the sum of all possible arm movements during the game state.
The root node starts with the rotation of the shoulder inside the XZ plane, or yaw, descending afterward to the second level where the decision resides in the degrees of rotation in the XY plane, or pitch. The third level is then reached, where the selection will determine the rotation of the arm around itself or roll. The final decision will determine the rotation of the elbow, which has only one degree of freedom. Having a complete path, from root to level four, which is a leaf, will determine the virtual position of the hand, which coincides with the desired position for the fruit placement. A path from the root node to a terminal node is then executed to reach that position. When descending the tree, the movement path of the avatar$'$s upper limb is determined by the distances between the position of a wrist, elbow and shoulder components. Traversing the tree means that avatar$'$s hand moves along the paths described by the current set of rotations until the final position is reached.
The general MCTS algorithm is depicted in ÒTable 1Ó, with a combination of paths selected during the selection and playout steps form a single simulation. Here $v_0$ is the root node of the initial state $s_0$, and $v_4$ is the last node reached during the descending stage through the tree and corresponds to the final position reached with the pseudo-random choices. The choices are not completely random as $BestChild$ will stimulate a certain degree of exploration, to avoid duplicate random unfolding of the game. $\Delta$ is the reward for the terminal state, $s_1$, reached by running the default algorithm from state $s_0$. This reward reflects the accuracy of the position reached, as the General MCTS is only capable of finding a "best solution" in a limited pool of samples.\\
\begin{table}[h]
\caption{Algorithm 1, General MCTS.}
\label{Table 1}
\begin{center}
\begin{tabular}{|c|}
\hline
\\
Function MonteCarloTreeSearch $(s_0 )$\\
Create root node $(v_0 )$  within state $(s_0 )$\\
$level$ $\leftarrow$ 0 \\
\\
While ($level$ \verb+<+ 4) \\
$v_{level + 1}$ $\leftarrow$ BestChild$(v_{level})$\\
$level$ $\leftarrow$ $level$ + 1 \\
End While \\
\\
$\Delta$ $\leftarrow$ Accuracy $(v_4 )$\\
BackPropagate $(\Delta)$\\
return $v_4$ \\
\\
\hline
\end{tabular}
\end{center}
\end{table}
As explained before, MCTS function has the capacity of looking ahead to obtain the accurate success probability of taking different paths (decision making). As for all real time games, the time constraint is very tight \cite{tong2010monte} expecting a player to make a decision towards an action within about $10$ $ms$ (milliseconds).\\
The next move is simulated based on the information taken from the current situation. Suppose that a fruit is located at $(x, y)$ we need to go along a path $d$ in order to collect it. The next object is located at  position $({x}\prime, {y}\prime)$ generated with an improved algorithm, related to the Upper Confidence bound for Tree, ÒEqn .2Ó. This encourages movements just outside of the player$'$s comfort zone. Each node $v$ in MCTS holds four pieces of data:
\begin{itemize}
\item Associated state $s(v)$ \item Incoming action $a(v)$ 
\item Total simulation reward $Q(v)$ 
\item Visit count $N(v)$ (a non negative integer) \cite{tong2010monte}
\\
\end{itemize}
The ratio of number of times the node has been visited Q(v), to the total reward of playouts N(v) is defined in Eqn .1,  \cite{browne2012survey}, \cite{tong2011monte}, \cite{pepels2014real} and \cite{tong2010monte}.
$$
\frac{Q(v)}{N(v)}  \eqno{(1)}
$$
The return or score of the overall search in the tree is $a(BestChild(v_0,0))$ with the action $a$ that leads to the child with successful results with the exploration parameter $c$. 
The score and time are backpropagated to $({x}\prime, {y}\prime)$ and another simulation from $(x, y)$ is started again. 
The mechanism of the MCTS algorithm is made of; selection, expansion, simulation and backpropagation \cite{chaslot2008monte} and \cite{browne2012survey}.\\

\begin{enumerate}[]
\item Selection: The algorithm starts at the root node, builds a tree of possible future game states (children). Each node in the tree represents a state and each link represents an action taken in that state that leads to a new state \cite{winands2008monte}. While the state is built in the tree, the next action is chosen according to the statistics stored, in a way that balances between exploitation and exploration. The task is either selecting an action that leads to the best results so far called exploitation or a less promising actions still needs to be explored known as; exploration, due to the uncertainty of the evaluation.
\begin{figure*}[!t]
\centering
\includegraphics[scale=0.4]{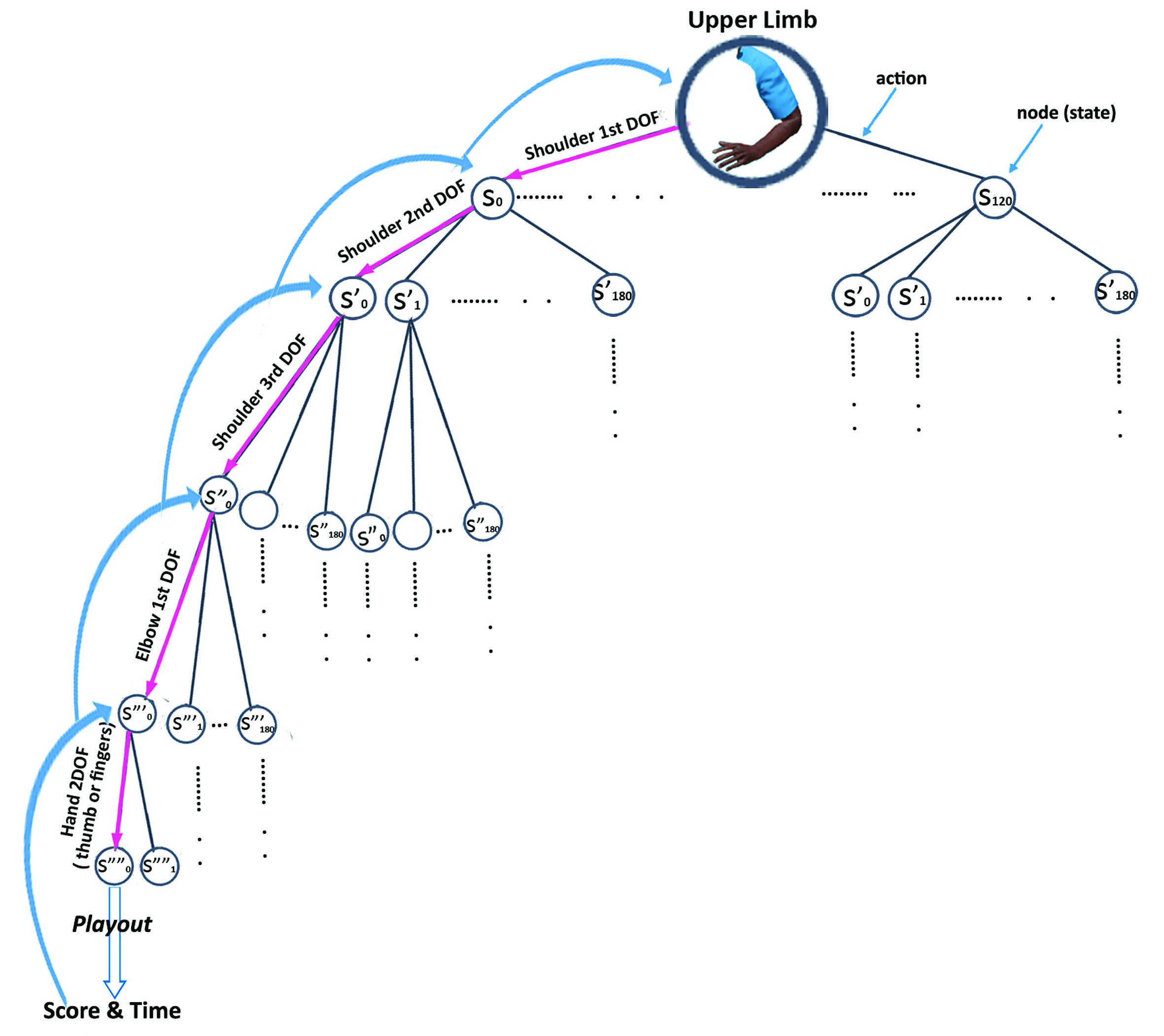}
%\label{fig_first_case}}
\hfil
\caption{Mechanism of Monte Carlo Tree Search algorithm of RehabGame.}
      \label{Fig. 4}
\end{figure*}

\item Expansion:  More state (child nodes) needs to be added as new nodes to expand the tree. The tree is expanded by one node for each simulated game according to the available actions. The first action is chosen by an expansion strategy and subsequently simulated. This results in a new game state, for which a new node is created. 
\item Simulation (Playout): After expansion, a rollout is done from the new node, which means that a simulated playout is run from the new node applying random actions until a predefined stop criterion is met or the game ends. The satisfactory weighting of action selection probabilities has a significant effect on the level of play. If all legal actions are selected with equal probability, then the strategy played is often weak, and the level of the Monte-Carlo program is suboptimal \cite{chaslot2008monte}.
\item Backpropagation (back up): The result of the simulated playout is propagated immediately from the rollout (selected node) back up to the root node. This means that the reward is saved to the visited nodes then a new iteration starts.
\end{enumerate}
Backpropagation updates node statistics that are utilized for future decision makings using Default Policy,  \cite{browne2012survey}. It playouts the domain from a given non-terminal state to produce a value estimation (simulation).

\subsection{Upper Confidence Bound for Tree}
The modified version of MCTS is an effective variant and selection strategy (UCB applied to trees algorithm-UCT) known as the upper confidence bound for trees. The UCT is derived from the UCB function for maximizing the rewards of a multi-armed bandit \cite{browne2012survey}, \cite{de2016monte}, \cite{ikehata2011monte} and \cite{frydenberg2015investigating}. UCT balances the choice between poorly explored actions with a high uncertainty and actions that have been explored extensively with a higher value with a child node $j$. If child nodes for all actions of the current node are added into the tree, actions are selected in order to maximize a UCT on the action value. UCT works by doing many multi-phase playouts:
$$
UCT=\bar X_j+C_p \sqrt \frac{\ln n}{n_j} \eqno{(2)}
$$

where $C_p$,  is a bias parameter which defines the proportion of exploitation and exploration. If $C_p= 0$, the UCB policy becomes a greedy policy \cite{vien2013monte}. $\bar X_j$ is the average reward from child $j$ and it is the exploitation part of the algorithm  and the second part is exploration part. $n$ is the number of times the current node (parent) has been played, $n_j$ ́  is the number of times child $j$ has been visited and $C_p>0$ is a constant (often set to $\surd2$) that shifts priority from exploration to exploitation. By increasing $C_p$, the priority is shifted to exploration which means; states that have been visited less will be visited with a higher priority than states that have been visited more. A decrease shifts the priority to exploitation and means that; states which have a high value are visited more in order to maximize reward. If more than one child node has the same maximal value, the tie is usually broken randomly. The values of $X_{i,t}$ and  $\bar X_j$ ̅are within $[0, 1]$. UCT value yields $\infty$, as per $n_j=0$ . The traditional expansion strategy is to explore each action at least once in each node. After all, actions have been expanded, the node applies the selection strategy for further exploration. After a rollout, the reward is backed up, which means that the estimated value for every node that has been visited in this iteration is updated with the reward of this simulation. Usually, the estimated value of a node is the average of all rewards backed up to that node. The UCT selection policy is used when the visit count of all children is above a threshold. When one or more of the children's visit counts are below this threshold, a random uniform selection is made. 

\subsection{Markov Decision Making process}
Markov Decision Processes (MDPs) are a fundamental modeling approach in decision theory and planning. They are used extensively within the artificial intelligence and operations research communities to model problems that require sequential decision making in an uncertain environment \cite{zhu2013real}. An MDP is generally defined by the subsequent elements: $S$ is a finite set of fully-observable possible states, $A$ is a finite set of possible actions depending on the states, $S\times A \times [0,1]$. It is the transition function where $P(s,a,s\prime )$ is the probability of moving to state $s\prime$ when action $a$ is applied in state $s$. $R$ is a real-valued reward function where $R(s,a)$ represents the expected reward for taking action $a$ in state $s$. $A$ policy $\pi$$:$$S\rightarrow A$ specifies an action $a = \pi(s)$ to be taken when in state $s$. The value function is defined as in Eqn.3:
$$
V_\pi (s)=R(s,a)+\gamma .\sum _{s\prime\in S} P(s,a,s\prime).V_\pi(s\prime)  \eqno{(3)}
$$

This represents the long term expected reward of executing a policy where $0$$\le$$\gamma$$\le$$1$ is a discount rate on future rewards. An optimal policy that maximizes the long term expected rewards is defined via the optimal value function shown in Eqn.4:
$$
V^\star (s)=max_{a\in A}R(s,a)+\gamma.\sum_{s\prime \in S} P(s,a,s\prime).V^\star (s\prime)  \eqno{(4)}
$$

\section{Monte Carlo Tree Search for RehabGame with decision-making process}
The main mechanism of the RehabGame resides in the fruit generator algorithm. This algorithm is an alternative variation of the UCT. The difference is that the statistical improvements that are calculated during the back-propagation phase, are also being propagated horizontally through the tree. This is achieved through the presence of virtual paths, or prospects. These prospects represent the nodes that have not been explored yet throughout the tree, which are grouped in layers. The statistical probabilities are added to these inexistent nodes using ÒEqn. 5Ó:
$$
 P_x = (x^2 + 1)^{-k}     \eqno{(5)}
$$
$P_x$ represents the prospect for the $x^{th}$ neighbor of the visited node, and $k$ represents the degree of propagation. The larger the value, the more localized the back-propagation is.\\
The Score is achieved depending on the avatar$'$s activity. If the fruit is collected and placed in a right place: $+1$. If the fruit is not reached or collected: $-1$. And if the fruit is collected but not placed in the basket or in the right location: $0$. 
$$
 M_{Score}^i= 
 \left\{
 \begin{array}{lll}
M_{success}^i &  if \ Score \ is \ +1\\
M_{unsuccess}^i &  if \ Score \ is \ 0\\
M_{fail}^i &  if \ Score \ is \ -1\\
                  
                \end{array}
              \right.\eqno{(6)}
$$ 
\begin{figure*}[thpb]
      \centering
     % \framebox{\parbox{30in}{}}
      \includegraphics[scale=1.0]{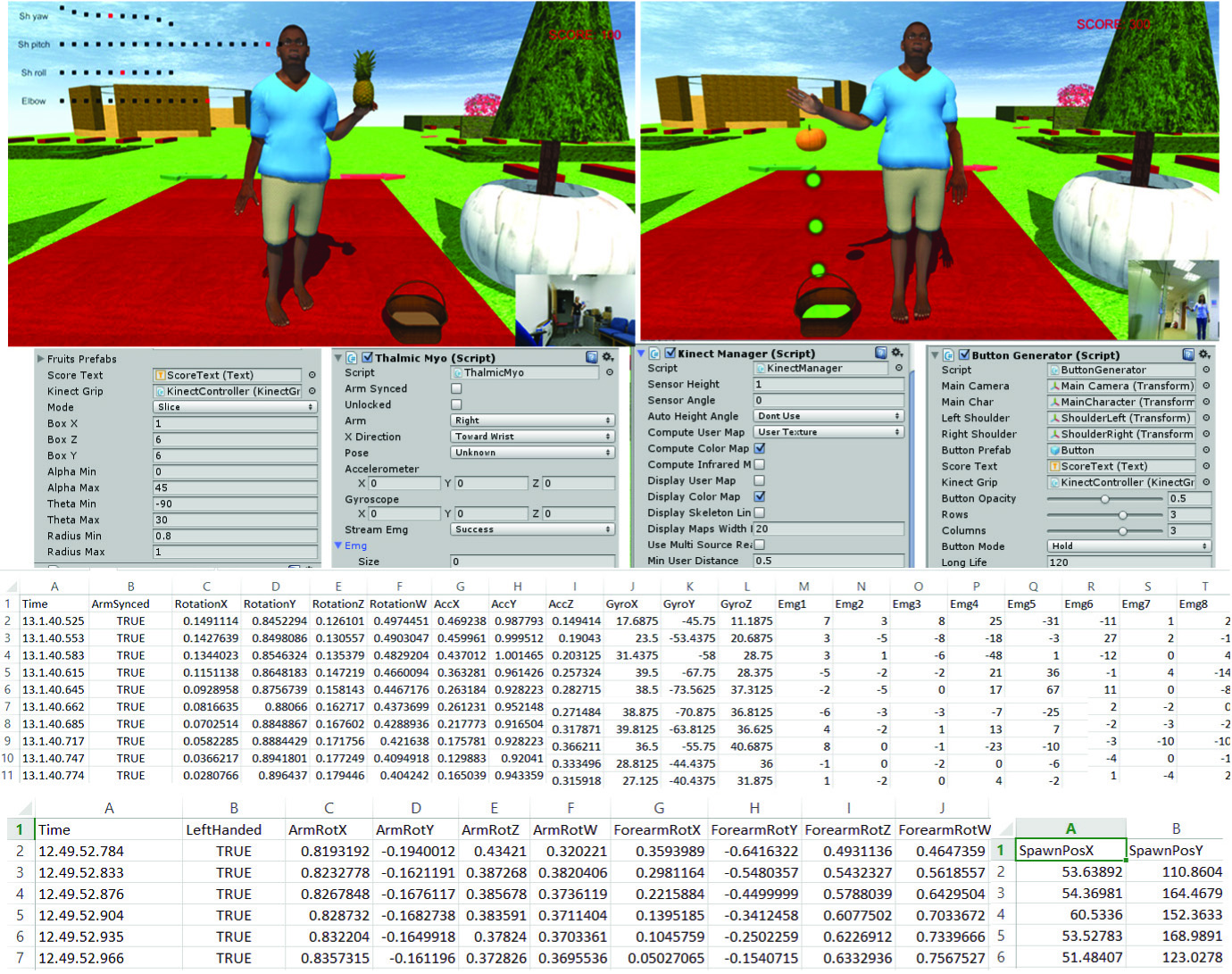}
      \caption{The configuration interface, parameters, results, constraints definition and Screen shots of RehabGame in which the avatar collects and releases the fruits above the basket.}
      \label{Fig. 5}
   \end{figure*}
$M_{Score}^i$ is the score of the $i^{th}$ play during the selection, backpropagation, and decision-making process as defined in ÒEqn. 6Ó. This Score is multiplied by the time efficiency, which is determined by confronting the player's time with a liner interpolation between a best possible time (max score) and a maximum time allowed (0 scores). Once the Score is determined, it is propagated upwards through the tree and expanded into the prospects of the neighbors, based on ÒEqn. 5Ó.\\
When a new fruit needs to be spawned the agent analyses and extracts information from the avatar$'$s past movements then selects the move with the most ambiguous success rate and based on that, it decides on the next step. By selecting the routes with the lowest absolute value, we are targeting the positions that have not been explored yet, or positions that a participant is training on, by migrating from achievable areas to unachievable areas.\\
In this process, the agent calculates the distances between avatar$'$s joints and the  fruits in the 3D scene using data collected from Kinect (position) and Myo (gyroscope). It generates the forthcoming fruits inside the player$'$s comfort zone and gradually directs it out of this zone while avoiding frustration or causing any harm. The distances are calculated as a vector from one joint to the other as well as from the joints to the fruit's location. The data generated will be saved after every session and can be retrieved by the agent and clinician. Fig.~\ref{Fig. 5} shows the interfaces and data generated in the background. 
The game can be set up for right or left upper limb, the orientation and muscle activity data in 3D space are accumulated in a file to be accessed by the agent and also for further investigation. The data produced are orientation and angle of rotations, muscles activity through $8$ EMGs, triaxial accelerometer and gyroscope that can detect inertial forces on all three axes every frame. Fig.~\ref{Fig. 6} shows the data taken from the sample while playing the game that illustrates EMG, accelerometer, gyroscope and EMGs data. This represents the muscle activity during the fruit pick up practice. 
\begin{figure}[thpb]
      \centering
     % \framebox{\parbox{30in}{}}
    \includegraphics[scale=0.8]{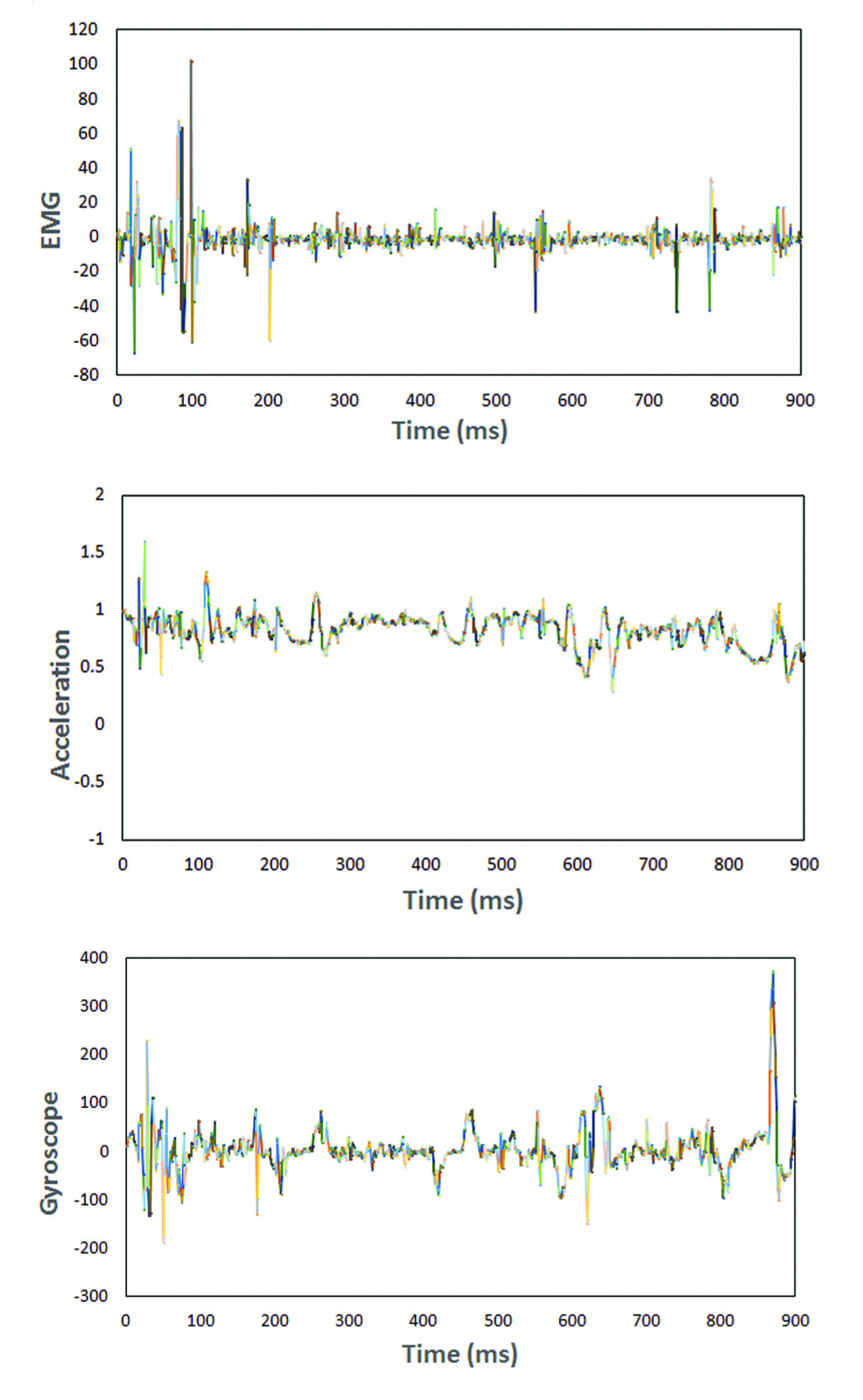}
      \caption{Results taken from Myo (Gyroscope, Accelerometer and EMG).}
      \label{Fig. 6}
   \end{figure}
The search path is variably determined by a distance limit $T_{path}$: $max_{P_1P_4}$.  According to \cite{adewusi2012biomechanical} the length of the upper limb is in average in the range of $665$ to $850$ mm with an average hand length of $184$ mm, forearm length of $260$ mm and upper arm length of $285$ mm.
The threshold defined in RehabGame is the max length of the arm. It ensures safe movements and paths are taken with high confidence in the safety and less damage to the muscle. Fruits are generated based on the paths created by the upper limb and the agent$'$s UCT simulation strategy based on various multi-phase playouts. 
A rewarding node in the tree might result in a loss at an attempt due to stochastic nature of the game. For that, a progressive expansion is performed for further exploration and exploitation of the tree.

\section{Conclusion}
RehabGame has the ability to modify and adapt its challenge level to match the skills of the player, avoiding frustration and encouraging engagement. It uses a variant of the Monte-Carlo Tree Search algorithm that involves the iterative building of search tree until some predefined computational constrains are achieved. At that point, the search is stopped and the best performing root action is returned. The current results may provide a starting point for other virtual reality studies and
rehabilitation programs compatible with cognitive impairment.
Future studies can use the current findings regarding differences in correlations as a way to generate
future hypotheses and guide research. The study was conducted with two sample sizes. Future research should
consider using a larger sample size, stroke, BI (brain injury) survivors and various ages.

\bibliographystyle{IEEEtran}

\bibliography{ref}

% Generated by IEEEtranS.bst, version: 1.14 (2015/08/26)
\begin{thebibliography}{10}
\providecommand{\url}[1]{#1}
\csname url@samestyle\endcsname
\providecommand{\newblock}{\relax}
\providecommand{\bibinfo}[2]{#2}
\providecommand{\BIBentrySTDinterwordspacing}{\spaceskip=0pt\relax}
\providecommand{\BIBentryALTinterwordstretchfactor}{4}
\providecommand{\BIBentryALTinterwordspacing}{\spaceskip=\fontdimen2\font plus
\BIBentryALTinterwordstretchfactor\fontdimen3\font minus
  \fontdimen4\font\relax}
\providecommand{\BIBforeignlanguage}[2]{{%
\expandafter\ifx\csname l@#1\endcsname\relax
\typeout{** WARNING: IEEEtranS.bst: No hyphenation pattern has been}%
\typeout{** loaded for the language `#1'. Using the pattern for}%
\typeout{** the default language instead.}%
\else
\language=\csname l@#1\endcsname
\fi
#2}}
\providecommand{\BIBdecl}{\relax}
\BIBdecl

\bibitem{adewusi2012biomechanical}
S.~Adewusi, S.~Rakheja, and P.~Marcotte, ``Biomechanical models of the human
  hand-arm to simulate distributed biodynamic responses for different
  postures,'' \emph{International Journal of Industrial Ergonomics}, vol.~42,
  no.~2, pp. 249--260, 2012.

\bibitem{billings2002challenge}
D.~Billings, A.~Davidson, J.~Schaeffer, and D.~Szafron, ``The challenge of
  poker,'' \emph{Artificial Intelligence}, vol. 134, no.~1, pp. 201--240, 2002.

\bibitem{browne2012survey}
C.~B. Browne, E.~Powley, D.~Whitehouse, S.~M. Lucas, P.~I. Cowling,
  P.~Rohlfshagen, S.~Tavener, D.~Perez, S.~Samothrakis, and S.~Colton, ``A
  survey of monte carlo tree search methods,'' \emph{IEEE Transactions on
  Computational Intelligence and AI in Games}, vol.~4, no.~1, pp. 1--43, 2012.

\bibitem{carlozzi2013using}
N.~E. Carlozzi, V.~Gade, A.~S. Rizzo, and D.~S. Tulsky, ``Using virtual reality
  driving simulators in persons with spinal cord injury: three screen display
  versus head mounted display,'' \emph{Disability and Rehabilitation: Assistive
  Technology}, vol.~8, no.~2, pp. 176--180, 2013.

\bibitem{chaslot2008monte}
G.~Chaslot, S.~Bakkes, I.~Szita, and P.~Spronck, ``Monte-carlo tree search: A
  new framework for game ai.'' in \emph{AIIDE}, 2008.

\bibitem{criminisi2011decision}
A.~Criminisi, J.~Shotton, and E.~Konukoglu, ``Decision forests for
  classification, regression, density estimation, manifold learning and
  semi-supervised learning,'' \emph{Microsoft Research Cambridge, Tech. Rep.
  MSRTR-2011-114}, vol.~5, no.~6, p.~12, 2011.

\bibitem{de2016monte}
M.~de~Waard, ``Monte carlo tree search with options for general video game
  playing,'' Ph.D. dissertation, Universiteit van Amsterdam, 2016.

\bibitem{dobkin2004strategies}
B.~H. Dobkin, ``Strategies for stroke rehabilitation,'' \emph{The Lancet
  Neurology}, vol.~3, no.~9, pp. 528--536, 2004.

\bibitem{finnsson2012generalized}
H.~Finnsson, ``Generalized monte-carlo tree search extensions for general game
  playing.'' in \emph{AAAI}, 2012.

\bibitem{frydenberg2015investigating}
F.~Frydenberg, K.~R. Andersen, S.~Risi, and J.~Togelius, ``Investigating mcts
  modifications in general video game playing,'' in \emph{2015 IEEE Conference
  on Computational Intelligence and Games (CIG)}.\hskip 1em plus 0.5em minus
  0.4em\relax IEEE, 2015, pp. 107--113.

\bibitem{holden2005virtual}
M.~K. Holden, ``Virtual environments for motor rehabilitation: review,''
  \emph{Cyberpsychology \& behavior}, vol.~8, no.~3, pp. 187--211, 2005.

\bibitem{holden2002virtual}
M.~K. Holden and T.~Dyar, ``Virtual environment training: a new tool for
  neurorehabilitation.'' \emph{Journal of Neurologic Physical Therapy},
  vol.~26, no.~2, pp. 62--71, 2002.

\bibitem{ikehata2011monte}
N.~Ikehata and T.~Ito, ``Monte-carlo tree search in ms. pac-man,'' in
  \emph{2011 IEEE Conference on Computational Intelligence and Games
  (CIG'11)}.\hskip 1em plus 0.5em minus 0.4em\relax IEEE, 2011, pp. 39--46.

\bibitem{levin2015emergence}
M.~F. Levin, P.~L. Weiss, and E.~A. Keshner, ``Emergence of virtual reality as
  a tool for upper limb rehabilitation: incorporation of motor control and
  motor learning principles,'' \emph{Physical therapy}, vol.~95, no.~3, pp.
  415--425, 2015.

\bibitem{lucas2015artificial}
S.~M. Lucas, M.~Mateas, M.~Preuss, P.~Spronck, and J.~Togelius, ``Artificial
  and computational intelligence in games: Integration (dagstuhl seminar
  15051),'' \emph{Dagstuhl Reports}, vol.~5, no.~1, 2015.

\bibitem{maccormick2011does}
J.~MacCormick, ``How does the kinect work?'' \emph{Presentert ved Dickinson
  College}, vol.~6, 2011.

\bibitem{maycock2013enhancing}
S.~Maycock and T.~Thompson, ``Enhancing touch-driven navigation using informed
  search in ms. pac-man,'' in \emph{Computational Intelligence in Games (CIG),
  2013 IEEE Conference on}.\hskip 1em plus 0.5em minus 0.4em\relax IEEE, 2013,
  pp. 1--2.

\bibitem{mirelman2009effects}
A.~Mirelman, P.~Bonato, and J.~E. Deutsch, ``Effects of training with a
  robot-virtual reality system compared with a robot alone on the gait of
  individuals after stroke,'' \emph{Stroke}, vol.~40, no.~1, pp. 169--174,
  2009.

\bibitem{mohanty2011teaching}
S.~D. Mohanty and S.~Cantu, ``Teaching introductory undergraduate physics using
  commercial video games,'' \emph{Physics Education}, vol.~46, no.~5, p. 570,
  2011.

\bibitem{nguyen2013monte}
K.~Q. Nguyen and R.~Thawonmas, ``Monte carlo tree search for collaboration
  control of ghosts in ms. pac-man,'' \emph{IEEE Transactions on Computational
  Intelligence and AI in Games}, vol.~5, no.~1, pp. 57--68, 2013.

\bibitem{o2013cognitive}
R.~L. O'Neil, R.~L. Skeel, and K.~I. Ustinova, ``Cognitive ability predicts
  motor learning on a virtual reality game in patients with tbi,''
  \emph{NeuroRehabilitation}, vol.~33, no.~4, pp. 667--680, 2013.

\bibitem{page2004efficacy}
S.~J. Page, S.~Sisto, P.~Levine, and R.~E. McGrath, ``Efficacy of modified
  constraint-induced movement therapy in chronic stroke: a single-blinded
  randomized controlled trial,'' \emph{Archives of physical medicine and
  rehabilitation}, vol.~85, no.~1, pp. 14--18, 2004.

\bibitem{pepels2014real}
T.~Pepels, M.~H. Winands, and M.~Lanctot, ``Real-time monte carlo tree search
  in ms pac-man,'' \emph{IEEE Transactions on Computational Intelligence and AI
  in Games}, vol.~6, no.~3, pp. 245--257, 2014.

\bibitem{pirovano2012self}
M.~Pirovano, R.~Mainetti, G.~Baud-Bovy, P.~L. Lanzi, and N.~A. Borghese,
  ``Self-adaptive games for rehabilitation at home,'' in \emph{2012 IEEE
  Conference on Computational Intelligence and Games (CIG)}.\hskip 1em plus
  0.5em minus 0.4em\relax IEEE, 2012, pp. 179--186.

\bibitem{saposnik2010effectiveness}
G.~Saposnik, R.~Teasell, M.~Mamdani, J.~Hall, W.~McIlroy, D.~Cheung, K.~E.
  Thorpe, L.~G. Cohen, M.~Bayley, S.~O. R. C. S.~W. Group \emph{et~al.},
  ``Effectiveness of virtual reality using wii gaming technology in stroke
  rehabilitation a pilot randomized clinical trial and proof of principle,''
  \emph{Stroke}, vol.~41, no.~7, pp. 1477--1484, 2010.

\bibitem{sheppard2002world}
B.~Sheppard, ``World-championship-caliber scrabble,'' \emph{Artificial
  Intelligence}, vol. 134, no.~1, pp. 241--275, 2002.

\bibitem{shotton2013real}
J.~Shotton, T.~Sharp, A.~Kipman, A.~Fitzgibbon, M.~Finocchio, A.~Blake,
  M.~Cook, and R.~Moore, ``Real-time human pose recognition in parts from
  single depth images,'' \emph{Communications of the ACM}, vol.~56, no.~1, pp.
  116--124, 2013.

\bibitem{smith1999task}
G.~V. Smith, K.~H. Silver, A.~P. Goldberg, and R.~F. Macko,
  ``“task-oriented” exercise improves hamstring strength and spastic
  reflexes in chronic stroke patients,'' \emph{Stroke}, vol.~30, no.~10, pp.
  2112--2118, 1999.

\bibitem{suma2013adapting}
E.~A. Suma, D.~M. Krum, B.~Lange, S.~Koenig, A.~Rizzo, and M.~Bolas, ``Adapting
  user interfaces for gestural interaction with the flexible action and
  articulated skeleton toolkit,'' \emph{Computers \& Graphics}, vol.~37, no.~3,
  pp. 193--201, 2013.

\bibitem{tan2009arm}
M.~P. Tan, N.~N. Wynn, M.~Umerov, A.~Henderson, A.~Gillham, S.~Junejo, and
  S.~K. Bansal, ``Arm span to height ratio is related to severity of dyspnea,
  reduced spirometry volumes, and right heart strain,'' \emph{CHEST Journal},
  vol. 135, no.~2, pp. 448--454, 2009.

\bibitem{tong2010monte}
B.~K. Tong and C.~W. Sung, ``A monte-carlo approach for ghost avoidance in the
  ms. pac-man game,'' in \emph{2010 2nd International IEEE Consumer Electronics
  Society's Games Innovations Conference}.\hskip 1em plus 0.5em minus
  0.4em\relax IEEE, 2010, pp. 1--8.

\bibitem{tong2011monte}
B.~K.-B. Tong, C.~M. Ma, and C.~W. Sung, ``A monte-carlo approach for the
  endgame of ms. pac-man,'' in \emph{2011 IEEE Conference on Computational
  Intelligence and Games (CIG'11)}.\hskip 1em plus 0.5em minus 0.4em\relax
  IEEE, 2011, pp. 9--15.

\bibitem{tsekleves2016development}
E.~Tsekleves, I.~T. Paraskevopoulos, A.~Warland, and C.~Kilbride, ``Development
  and preliminary evaluation of a novel low cost vr-based upper limb stroke
  rehabilitation platform using wii technology,'' \emph{Disability and
  Rehabilitation: Assistive Technology}, vol.~11, no.~5, pp. 413--422, 2016.

\bibitem{ustinova2014virtual}
K.~Ustinova, J.~Perkins, W.~Leonard, and C.~Hausbeck, ``Virtual reality
  game-based therapy for treatment of postural and co-ordination abnormalities
  secondary to tbi: A pilot study,'' \emph{Brain injury}, vol.~28, no.~4, pp.
  486--495, 2014.

\bibitem{vien2013monte}
N.~A. Vien, W.~Ertel, V.-H. Dang, and T.~Chung, ``Monte-carlo tree search for
  bayesian reinforcement learning,'' \emph{Applied intelligence}, vol.~39,
  no.~2, pp. 345--353, 2013.

\bibitem{villiger2012virtual}
M.~Villiger, ``Virtual reality rehabilitation in spinal cord injury patients,''
  Ph.D. dissertation, Diss., Eidgen{\"o}ssische Technische Hochschule ETH
  Z{\"u}rich, Nr. 20482, 2012, 2012.

\bibitem{villiger2011virtual}
M.~Villiger, M.-C. Hepp-Reymond, P.~Pyk, D.~Kiper, K.~Eng, J.~Spillman,
  B.~Meilick, N.~Est{\'e}vez, S.~S. Kollias, A.~Curt \emph{et~al.}, ``Virtual
  reality rehabilitation system for neuropathic pain and motor dysfunction in
  spinal cord injury patients,'' in \emph{2011 International Conference on
  Virtual Rehabilitation}.\hskip 1em plus 0.5em minus 0.4em\relax IEEE, 2011,
  pp. 1--4.

\bibitem{weiss2013video}
P.~L. Weiss, H.~Sveistrup, D.~Rand, and R.~Kizony, ``Video capture virtual
  reality: a decade of rehabilitation assessment and intervention,''
  \emph{Physical Therapy Reviews}, 2013.

\bibitem{whitehouse2014monte}
D.~Whitehouse, ``Monte carlo tree search for games with hidden information and
  uncertainty,'' Ph.D. dissertation, University of York, 2014.

\bibitem{winands2008monte}
M.~H. Winands, Y.~Bj{\"o}rnsson, and J.-T. Saito, ``Monte-carlo tree search
  solver,'' in \emph{International Conference on Computers and Games}.\hskip
  1em plus 0.5em minus 0.4em\relax Springer, 2008, pp. 25--36.

\bibitem{zhu2013real}
G.~Zhu, ``Real-time elective admissions planning for health care providers,''
  2013.

\end{thebibliography}

\ifCLASSOPTIONcaptionsoff
  \newpage
\fi

%\begin{IEEEbiography}
%[{\includegraphics[width=1in,height=1.25in,clip,keepaspectratio]{fig6}}]{Shabnam Sadeghi Esfahlani}
%[{\includegraphics[clip,keepaspectratio]{mypic}}]{Michael Shell}
% or if you just want to reserve a space for a photo:
%Shabnam is a senior lecturer in Computer Gaming Technology, Anglia Ruskin University, Cambridge, UK. She received her Ph.D. in 2012 in Mechanical Engineering. She has research experience in virtual reality based rehabilitation game, model reliability, robustness and design optimization using applied Finite Element Analysis and statistical models. Her research interests include development of edutainment games for adults, and high-performance energy absorbing materials and meta-models.
%\end{IEEEbiography}

\end{document}